\documentclass[aps,twocolumn,tightenlines,amsmath,amssymb,superscriptaddress]{revtex4-2}

\usepackage{graphicx}
\usepackage{amsmath}
\usepackage{amssymb}
\usepackage{epstopdf}
\usepackage{color}
\usepackage{lipsum}

\usepackage[colorlinks=true]{hyperref}
\hypersetup{citecolor=blue}

\usepackage{ulem}
\usepackage[dvipsnames]{xcolor}
\usepackage{todonotes}
\DeclareGraphicsExtensions{.eps}

\begin{document}
\title{Single-active-element demultiplexed multi-photon source}
\author{L.~M.~Hansen}
\email{lena.maria.hansen@univie.ac.at}
\affiliation{University of Vienna, Faculty of Physics, Vienna Center for Quantum Science and Technology (VCQ), 1090 Vienna, Austria}
\affiliation{Christian Doppler Laboratory for Photonic Quantum Computer, Faculty of Physics, University of Vienna, Vienna, Austria}
\affiliation{Institut für Festkörperphysik, Technische Universität Berlin, 10623 Berlin, Germany.}
\author{L.~Carosini}
\affiliation{University of Vienna, Faculty of Physics, Vienna Center for Quantum Science and Technology (VCQ), 1090 Vienna, Austria}
\affiliation{Christian Doppler Laboratory for Photonic Quantum Computer, Faculty of Physics, University of Vienna, Vienna, Austria}
\author{L.~Jehle}
\affiliation{University of Vienna, Faculty of Physics, Vienna Center for Quantum Science and Technology (VCQ), 1090 Vienna, Austria}
\author{F.~Giorgino}
\affiliation{University of Vienna, Faculty of Physics, Vienna Center for Quantum Science and Technology (VCQ), 1090 Vienna, Austria}
\affiliation{Christian Doppler Laboratory for Photonic Quantum Computer, Faculty of Physics, University of Vienna, Vienna, Austria}
\author{R.~Houvenaghel}
\affiliation{Physics Department, Ecole Normale Supérieure de Lyon, 46 allée d’Italie, F69007 Lyon, France}
\author{M.~Vyvlecka}
\affiliation{University of Vienna, Faculty of Physics, Vienna Center for Quantum Science and Technology (VCQ), 1090 Vienna, Austria}
\author{J.~C.~Loredo}
\email{juan.loredo@univie.ac.at}
\affiliation{University of Vienna, Faculty of Physics, Vienna Center for Quantum Science and Technology (VCQ), 1090 Vienna, Austria}
\affiliation{Christian Doppler Laboratory for Photonic Quantum Computer, Faculty of Physics, University of Vienna, Vienna, Austria}
\author{P.~Walther}
\affiliation{University of Vienna, Faculty of Physics, Vienna Center for Quantum Science and Technology (VCQ), 1090 Vienna, Austria}
\affiliation{Christian Doppler Laboratory for Photonic Quantum Computer, Faculty of Physics, University of Vienna, Vienna, Austria}

\begin{abstract}
Temporal-to-spatial demultiplexing routes non-simultaneous events of the same spatial mode to distinct output trajectories. This technique has now been widely adopted because it gives access to higher-number multi-photon states when exploiting solid-state quantum emitters. However, implementations so far have required an always-increasing number of active elements, rapidly facing resource constraints. Here, we propose and demonstrate a demultiplexing approach that utilizes only a single active element for routing to, in principle, an arbitrary number of outputs. We employ our device in combination with a high-efficiency quantum dot based single-photon source, and measure up to eight demultiplexed highly indistinguishable single photons. We discuss the practical limitations of our approach, and describe in which conditions it can be used to demultiplex, e.g., tens of outputs. Our results thus provides a path for the preparation of resource-efficient larger-scale multi-photon sources.
\end{abstract}
\maketitle

\section{Introduction}

Advances in photonic quantum science~\cite{optqc:2007,PQTech:OBrien09,PQSim:Aspuru12,Flamini_2018,PQIP_Pryde_2019} occur with increased complexity of the available sources of non-classical light. The main technologies to date for producing single-, and multi-photon states are based on either frequency-conversion in non-linear media~\cite{Weston:16,Spring:17,Graffitti:18}, or atomic transitions of quantum emitters~\cite{sps:senellart17,UPPU:20,Tomm:2021aa}. The former produces heralded single-, and entangled-photon statistics, or squeezed states of light, and the latter primarily results in deterministic single-photon emission at high efficiencies and rates. The most advanced multi-photon experiments thus far involved the interference of up to 14 particles using single-photon sources~\cite{12photon_2018,BS_20photon_2019}, or the detection of hundreds of photons using squeezed light sources~\cite{QAdv_photon20,Madsen:2022aa}.

Temporal-to-spatial demultiplexing has played a key role in enabling these levels of complexity. This technique allowed the preparation of the multi-photon sources in the cases of space-encoded interference~\cite{BS_20photon_2019}, and enabled the measurement of---up to 16---consecutive temporal modes from time-bin interferometers~\cite{Madsen:2022aa}. This protocol deals with routing subsequent events, or time bins, from one spatial mode towards different locations, and it has been widely used to allow multi-photon experiments using quantum emitters~\cite{lenzini:17,Anton:19,Hummel.2019,BS_20photon_2019,Munzberg.2022,demux:Crespi22}. Indeed, creating multiple indistinguishable photons from a single demultiplexed source is still technically more viable than fabricating many quantum emitters that produce indistinguishable photons, where the state-of-the-art remains in the demonstration of photon indistinguishability from two remote sources~\cite{TPI:Warburton22}.

The standard approach for building a temporal-to-spatial demultiplexer starts by using an active element---e.g., an electro-optic modulator (EOM)---for producing orthogonal polarizations in two subsequent photons, later following different trajectories after traversing a polarization selective element, such as a polarizing beam splitter (PBS). By repeating this process at each new output, any number $N$ of consecutive time bins can be demultiplexed, however at the increasing cost of using $N{-}1$ active elements. At the output of the demultiplexer, photons originally separated in units of a temporal distance $\tau$ are appropriately delayed such that they all travel simultaneously and can interfere. Most implementations to date have employed this method~\cite{lenzini:17,Anton:19,Hummel.2019,Munzberg.2022}, with the current record of using $19$ high-voltage bulk EOMs for obtaining 20 photons~\cite{BS_20photon_2019}, each occupying one spatial mode. Evidently, this approach becomes resource expensive and costly.

Here, we demonstrate a temporal-to-spatial demultiplexer that uses only one active element to produce, in principle, an arbitrary number of outputs. We combine our device with a highly-efficient single-photon source from a quantum dot, and demonstrate the generation of an eight-photon state, where each indistinguishable photon is propagating in a separate spatial mode. Our scheme can be extended to bigger multi-photon states, however, in our case we are limited by efficiency of the photon source. One interesting feature of our approach is that the demultiplexed time bins can be as close as a few nanoseconds apart, which is beneficial for maintaining high levels of photon indistinguishability from quantum emitters~\cite{scalable:loredo16,transformLimit:JWP16}. Our implementation underlines the feasibility of our scheme that can enable near future practical multi-photon sources at larger scale.

	\begin{figure*}[htp!]
		\centering
		\includegraphics[width=.85\textwidth]{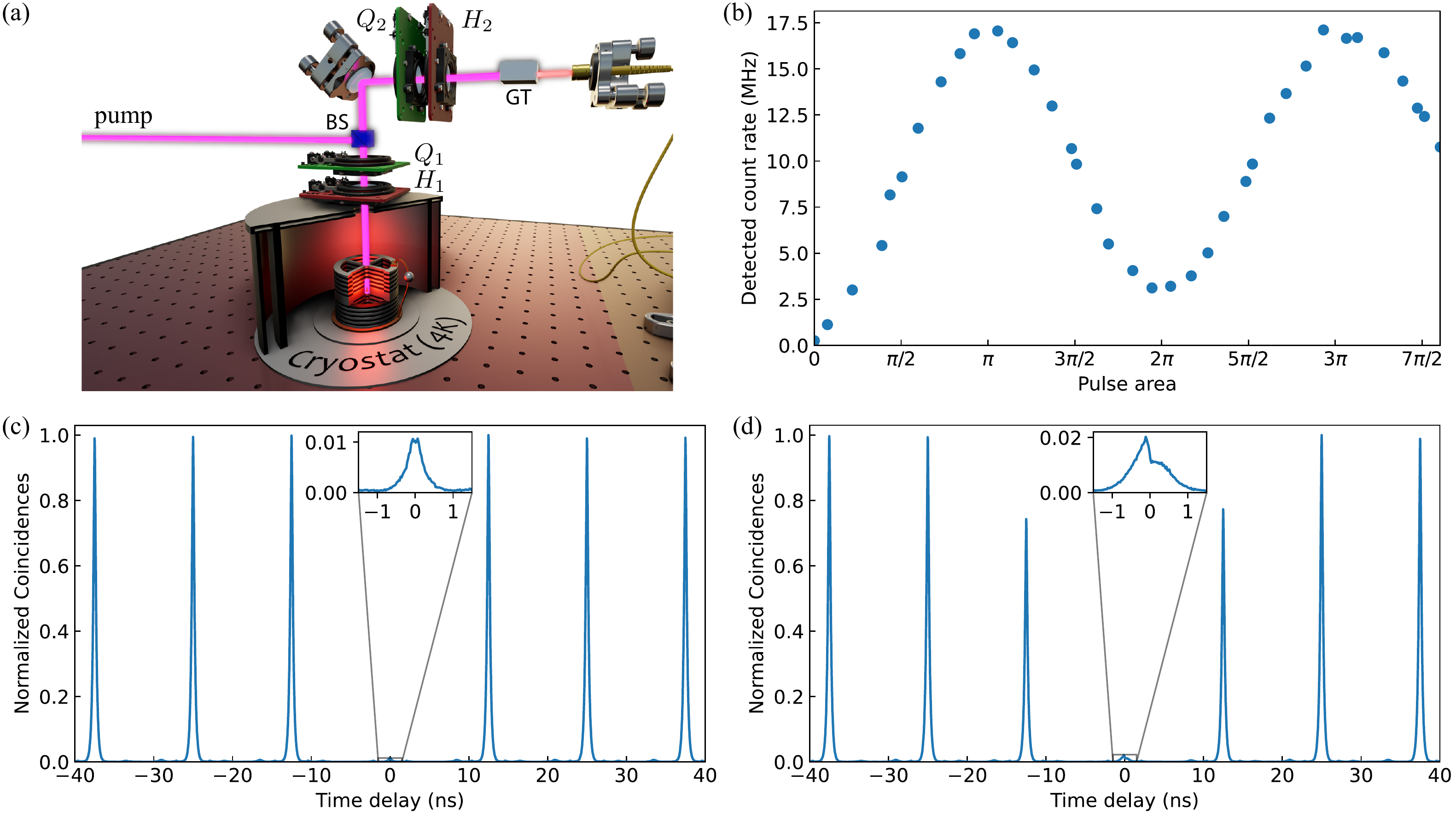}
		\caption{ \textbf{Single-photon source.} (a) Schematic figure of the excitation and collection setup with crossed-polarized configuration to separate pump light from single-photon emission. (b) Rabi oscillations: detected single-photon countrate vs pump pulse area. (c) Second-order auto-correlation measurement at $\pi$-pulse excitation. The value at zero time delay is $g^{(2)}(0){=}{1.57(2)}{\%}$. (d) Hong-Ou-Mandel interference at $\pi$-pulse, resulting in photon indistinguishability $\mathcal{I}{=}{95.35(3)}{\%}$. Both values in (c) and (d) are obtained by integrating peak areas in a 3~ns window. Measurement uncertainties are estimated following Poissonian counting statistics.}
	\label{fig:1}
	\end{figure*}

\newpage
	
\section{Source}	
A sample containing semiconductor quantum dots (QDs) in micropillar cavities is placed inside a cryostat at $\sim$4K, see Fig.~\ref{fig:1}(a). A QD is resonantly driven with laser pulses of 80~MHz repetition rate, and it is spectrally tailored to match the QD cavity wavelength of {922.2}~nm, and linewidth at FWHM of $~120$~pm. We use a $97{:}3$ beam splitter (BS) to guide a fraction of the laser pump towards the sample, while maintaining high transmission for the emitted single photons. The pump pulse is then focused on the sample by an aspheric lens placed inside the cryostat. We use a cross-polarized configuration for optical excitation and collection. In the excitation path, we control the power of the laser light, and initialize its polarization with a quarter-, and a half-wave plate, $Q_1,H_1$, respectively, to one of the linear-polarization cavity modes. In the collection path, we place another set of wave plates, $Q_2,H_2$, and a Glan-Taylor polarizer (GT) to suppress the laser by more than seven orders-of-magnitude. Thus, only single photons are coupled into the fiber of the collection setup.

	\begin{figure*}[htp!]
		\centering
		\includegraphics[width=0.95\textwidth]{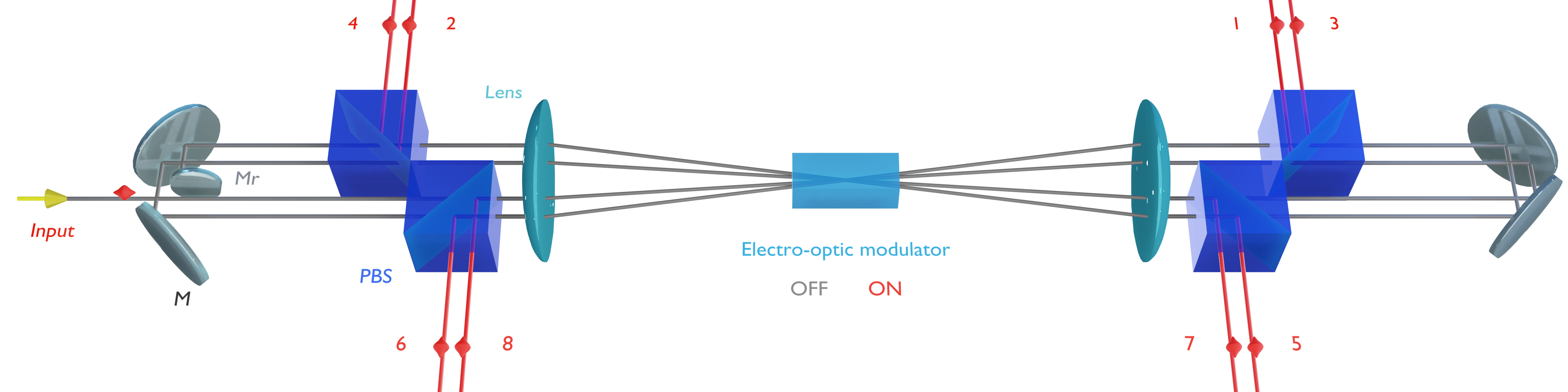}\vspace{0mm}
		\caption{\textbf{Demultiplexer setup.} A stream of single photons emitted by the QD enters the demultiplexer with horizontal polarization. During the loading phase, polarization is maintained and photons follow the gray-colored trajectory. The temporal delay introduced by the free-space path matches the temporal separation of the single photons.  For the release phase, the EOM switches the photons' polarization to vertical, such that they are reflected at the PBSs. The input stream that initially contained $N$ single photons in the same spatial mode and in consecutive time bins, is transformed into $N$ distinct trajectories containing a single photon each, in the same time bin.}
	\label{fig:2}
	\end{figure*}	

	\begin{figure*}[htp!]
		\centering
		\includegraphics[width=0.9\textwidth]{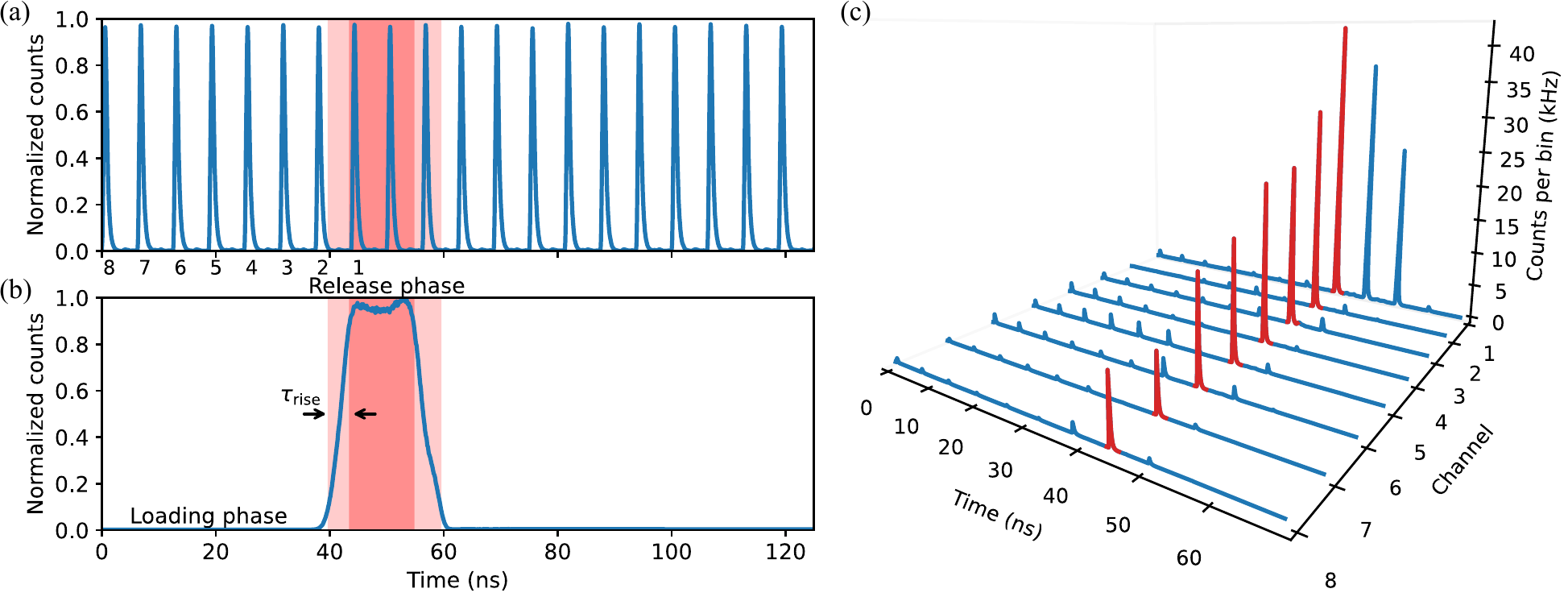}	\vspace{0mm}
		\caption{\textbf{Time traces.} (a) Single-photon stream generated at 160~MHz. The red colored area indicates the on state of the EOM. The numbering denotes the demultiplexer channel at which they exit. (b) Time modulation of the EOM. The loading phase has a duration of about 40~ns, and a switching time (light-red colored area) of $\sim$5~ns to the release phase. Approximately 25~ns pass before the EOM returns to the off state. (c) Time trace of all demultiplexer outputs. Eight single photons (red peaks) are demultiplexed into different spatial modes in the same time bin. The intensity of the correctly routed peak of a given channel is decreased by the sum of incorrectly routed peaks appearing in the loading phase among all previous channels.}
	\label{fig:3}
	\end{figure*}	
	
We use this setup to characterize our source. Figure~\ref{fig:1}(b) displays Rabi oscillations as a signature of the coherent driving of the system. At $\pi$-pulse excitation, we measure a maximum single-photon count rate of $17.1$~MHz---that is, an end-to-end source efficiency of $21.4{\%}$---recorded with a superconducting nanowire single-photon detector of $85{\%}$ system efficiency. We now characterize the single-photon purity by measuring the second-order auto-correlation function at zero time delay $g^{(2)}(0)$ using a standard Hanbury Brown and Twiss setup. At $\pi$-pulse excitation, we retrieve a value of $g^{(2)}(0){=}{1.57(2)}{\%}$, as shown in Fig.~\ref{fig:1}(c). At these same conditions, a two-photon Hong-Ou-Mandel (HOM) interference experiment~\cite{HOM87} reveals a photon indistinguishability~\cite{HOM:Ollivier21} $\mathcal{I}{=}95.35(3){\%}$, see Fig.~\ref{fig:1}(d). Moreover, lifetime measurements reveal a single-photon decay rate of $207.4(5)$~ps, as shown in the Supplementary Material.

\section{Single-active-element Demultiplexer}

The second and main part of our experiment consists of the temporal-to-spatial demultiplexer. It employs a single electro-optical modulator placed at the center of a near-recurrent geometry, as shown in Fig.~\ref{fig:2}. The temporal sequence of single photons emitted by the QD is guided to the demultiplexing setup via single-mode fibers, here the photons' polarization is set to horizontal. The scheme consists of two phases, a loading phase, and a release phase. During the loading part, the EOM is off, maintaining horizontal polarization. Here, photons follow trajectories slightly displaced from the center of a telescope of unity magnification made of two converging lenses, with the EOM placed in its center. After the PBS transmits the horizontal modes, the single photons follow a free-space delay line, matched to the separation between the input photons, and reenter the setup on a new parallel trajectory. This process is repeated several times, until a mirror $M_r$ back-reflects the photons' trajectories doubling the number of photons that pass through the EOM. For the release phase, with all modes loaded into the setup, we turn on the EOM, switching the photons' polarization to vertical. Now the PBSs reflect the photons simultaneously into distinct spatial output modes. This geometry can in principle continue to increase the number of optical paths as long as the optical elements' clear apertures allow it. In practice, the initial single-photon source efficiency is one main limiting factor for building multi-photon sources, as it determines the exponentially decreasing multi-photon rates. In this work, our source efficiency enables us to demultiplex up to eight single-photons.

We now operate and characterize our demultiplexer with the input single-photon source of Fig.~\ref{fig:1}. First, we insert photons separated by a shorter delay of $\tau{=}{6.25}$~ns, see Fig.~\ref{fig:3}(a), for which we passively increase the repetition rate of our laser~\cite{MultiPulse:Broome11} to $f_{\mathrm{L}}{=}\tau^{-1}{=}{160}$~MHz. We use this reduced temporal distance to allow for a shorter free-space delay. We drive the EOM with a frequency of $f_{\mathrm{EOM}}{=}{8}$~MHz phase-locked to the laser, a switching rise time of $\sim$5 ns, and an on-state duration of $25$~ns, see Fig.~\ref{fig:3}(b). Note that although EOMs with faster repetition rates exists to date, reducing the modulation rise time is favored in the present architecture.

For as long as the EOM is off---that is, during the loading phase---ideally no signal should be observed at any demultiplexer output. Thereon, as a result of the synchronized modulation on the input single-photon stream, every time the EOM reaches its on state---every 125~ns---a number of accumulated single photons is released, ideally one photon at every demultiplexer output. Figure~\ref{fig:3}(c) displays such time traces for eight outputs.

In an ideal implementation, we expect peaks only appearing in the targeted spatial-temporal outputs, and repeated every 125~ns, while no signal should be observed at all other times. In practice, we find additional peaks in other time bins. Here, we distinguish between three cases native to our architecture. In the first case, additional prominent peaks appear in the first output channel. They exist because the modulation pulse-width is longer than the separation between time bins. Thus, single photons that continue entering the setup after the start of the release phase are directly switched and reflected by the second PBS into the first output channel. However, these events have no impact in the resulting $N$-photon rates, and if necessary they can be suppressed prior to entering the demultiplexer. In a second case, incorrect routing of events occurs during the loading process, showing that the horizontal polarization of every path is not optimally maintained, such that photons reflect with a small probability when passing through the PBSs. This occurs because the trajectories followed by the photons are slightly misaligned from normal incidence towards the EOM central axis. This second case is a main reason for performance degradation in this scheme. Albeit small, the probability of incorrectly releasing a photon at any time bin of the loading phase accumulates, and thus it increases with higher number of outputs. In the third and final case, the EOM imperfectly switches the photons' polarization during the release phase, i.e., the maximum probability of reflecting is smaller than unity. Thus, some photons continue on the delay path and are likely to be released at the next output channel in the following time bin.

To assess the performance of the single-active-element demultiplexer, we now estimate the channel efficiency $\eta_{\text{ch}}$ of every output. This denotes the probability of a given input time bin---labeled from 1 to 8 in Fig.~\ref{fig:3}(a)---to be released in the targeted output---red-colored peaks in Fig.~\ref{fig:3}(c). We obtain efficiencies ranging from {$\eta_{\text{ch}1}{=}0.73$ for the first channel, to $\eta_{\text{ch}8}{=}0.14$ for the eighth channel,} see Supplementary Material for the list of channel efficiencies. These values are estimated as the ratio of the integrated counts of the main peak in a given output channel, to the counts of the corresponding input time bin. Note that this efficiency parameter is affected by any reason that lead to a reduced performance---for instance, losses of the optical elements, fiber-coupling losses, losses in connecting and mating fibers, as well as incorrect active switching of the EOM---this being the main contributing factor.

	\begin{figure}[htp!]
		\centering
		\includegraphics[width=0.48\textwidth]{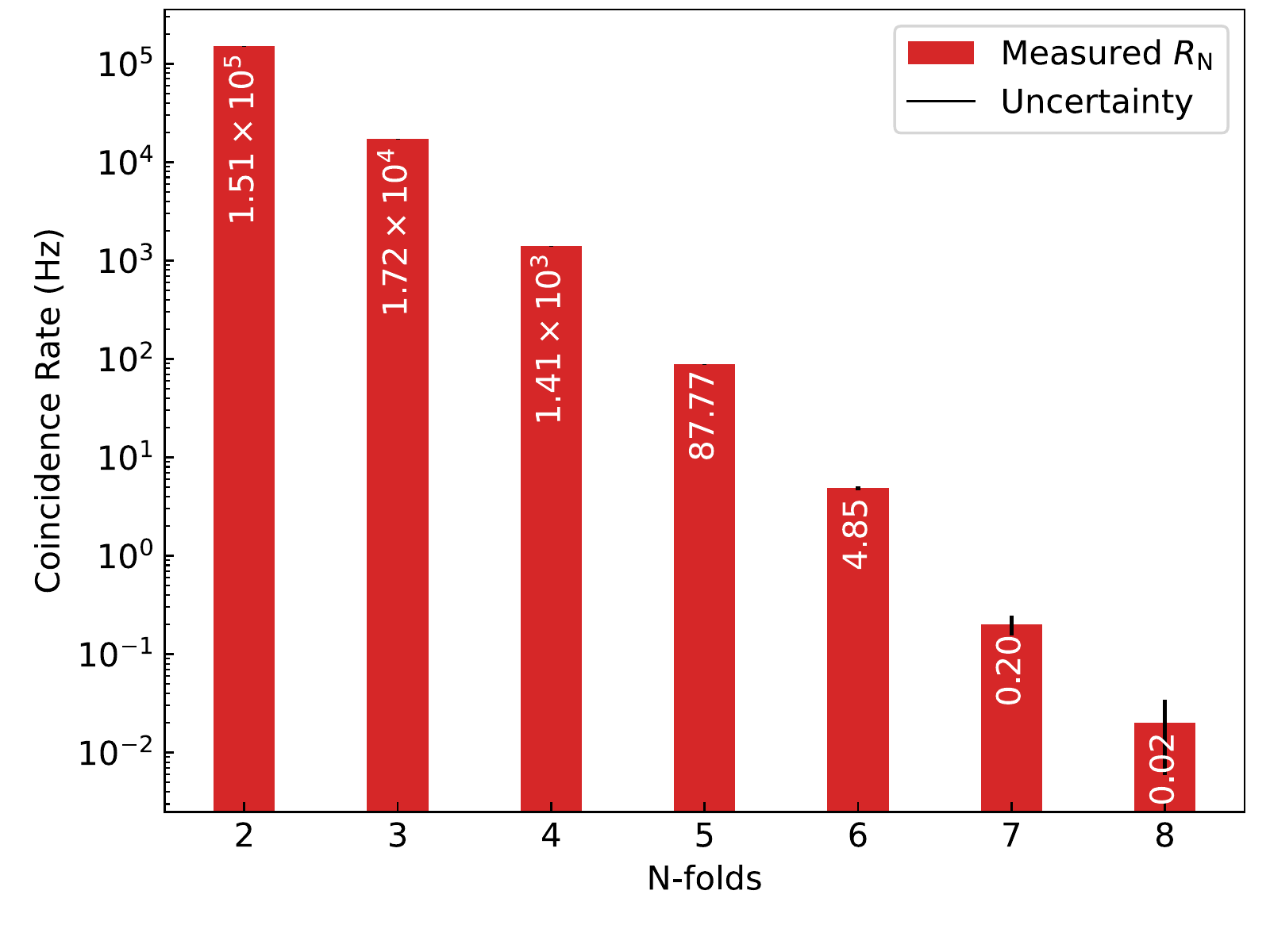}	\vspace{-10mm}
		\caption{\textbf{Multi-photon rates.} Measured multi-fold coincidence rates $R_N$: simultaneous output $N$-photon events detected per second. We integrate over a coincidence window of 3~ns. Uncertainties are obtained from Poissonian counting statistics, and are too small to be visible for smaller $N$.}
	\label{fig:4}
	\end{figure}

We now report the main figure-of-merit of our demultiplexing scheme: the measured multi-fold coincidence rate $R_N$---that is, the simultaneous detection of $N$ photons in distinct output modes, see Fig.~\ref{fig:4}. Here, the four-photon coincidence rate allows a direct comparison of our results to previous works. In our case, we detect a four-photon rate of ${\sim}$1.4~kHz, a minimum of three orders-of-magnitude higher than almost all previous implementations, with the exception of Ref.~\cite{BS_20photon_2019}, where our work compares similarly in the same count rate scale. The values $R_N$ depend and decrease exponentially with the source efficiency, as well as channel efficiencies. In our work, the high levels of efficiencies enable us to measure up to eight photons at a coincidence rate of ${\sim}20$~mHz. Our implementation, therefore, is at the state-of-the-art of active spatial-to-temporal demultiplexing schemes, but notably by using only a single active element.


\section{Discussion and conclusion}

In this work we presented a resource-efficient scheme for temporal-to-spatial mode demultiplexing, and use it in combination with a highly-efficient quantum dot based single-photon source. Our architecture requires only one active element for obtaining, in principle, an arbitrary number of outputs, thus significantly reducing the amount of resources compared to former implementations.  At present, our demultiplexer enables up to eight-photon coincidence events measured at a rate of ${\sim}20$~{mHz}. This constitutes a significant improvement in number of output channels and count rates compared to alternatives, locating our work at the state-of-the-art for multi-photon sources. 

Moreover, count rates following this approach can still significantly improve---for the same source efficiency---by addressing the factors affecting the channel efficiencies. The main limiting factor is the leakage of photon events in incorrect time bins during the loading process. This is due to the diagonal propagation of the single-photon modes relative to the central axes of the birefringent crystal, causing a small but accumulative polarization walk-off. While individually controlling the polarization of each path entering the EOM mitigates this effect to some extent, optimized channel efficiencies will require making use of compensation crystals, or modified trajectories that exploit near-recurrent geometries while imposing normal incidence along the birefringent element. Furthermore, reducing the path delay---consequently, using an EOM with faster rise time---within the demultiplexer is largely beneficial: smaller distances allows for smaller beam diameters and Rayleigh ranges, therefore many more trajectories can fit within limited optics' clear apertures. As proof-of-principle, we also built a demultiplexer unit that outputs sixteen modes using standard one inch optical elements, see Supplementary Material for more information. With these modifications in mind, our approach can be used to build multi-photon sources at larger scales.\\

{\bf Funding.}~This research was funded in whole, or in part, from the European Union’s Horizon 2020 and Horizon Europe research and innovation programme under grant agreement No 899368 (EPIQUS), the Marie Skłodowska-Curie grant agreement No 956071 (AppQInfo), and the QuantERA II Programme under Grant Agreement No 101017733 (PhoMemtor); from the Austrian Science Fund (FWF) through [F7113] (BeyondC), and [FG5] (Research Group 5); from the Austrian Federal Ministry for Digital and Economic Affairs, the National Foundation for Research, Technology and Development and the Christian Doppler Research Association. For the purpose of open access, the author has applied a CC BY public copyright licence to any Author Accepted Manuscript version arising from this submission.\\

{\bf Acknowledgment.}~The authors thank Patrik Zah\'alka for assistance with FPGA electronics and signal processing.\\

\bibliography{biblio_demux}
\break
\section{Supplementary Material}

\subsection{Lifetime}
	\begin{figure}[htp!]
		\centering
		\includegraphics[width=0.43\textwidth]{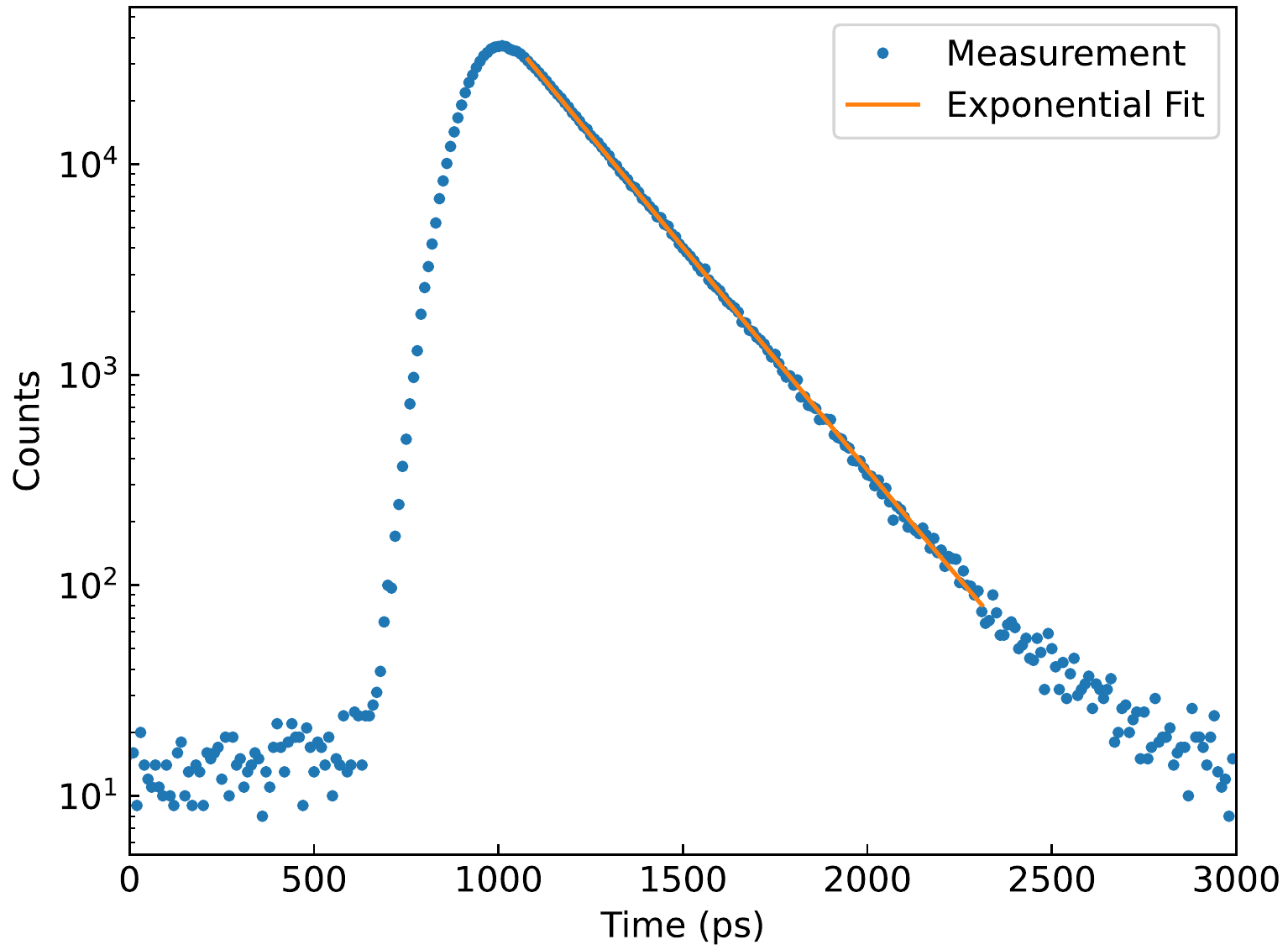}\vspace{0mm}
		\caption{\textbf{Single-photon lifetime.} Measurement of the decay dynamics of the QD at $\pi$-pulse excitation. From this measurement, we estimate a lifetime of {207.4(5)}~ps by fitting a mono-exponential decay.}
	\label{figSM:1}
	\end{figure}	

\subsection{Channel efficiencies}
	\begin{figure}[htp!]
		\centering
		\includegraphics[width=0.43\textwidth]{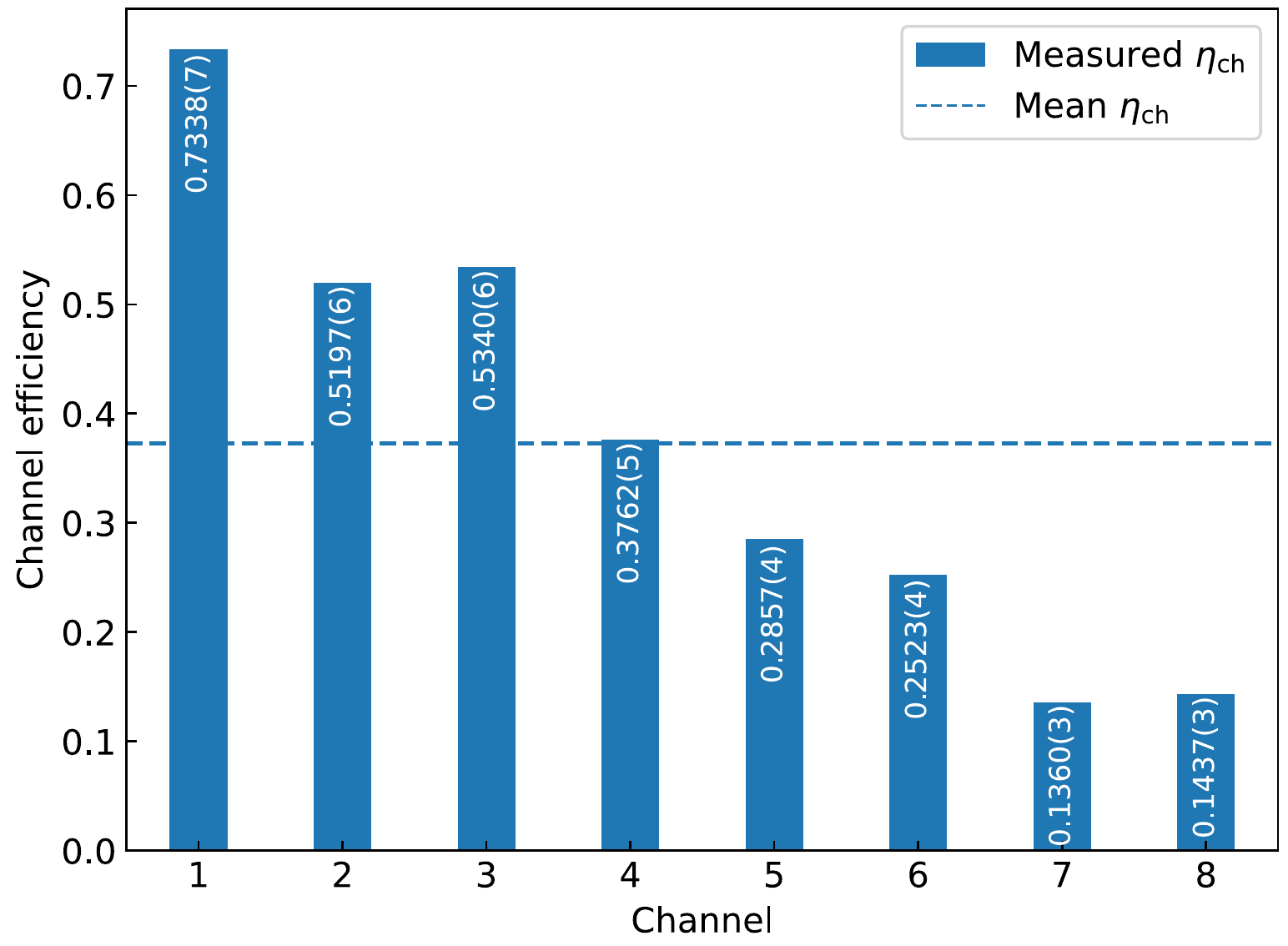}\vspace{0mm}
		\caption{\textbf{Channel efficiencies.} List of efficiencies for all eight demultiplexed outputs.}
	\label{figSM:2}
	\end{figure}	

\subsection{16-output demultiplexer}
	\begin{figure*}[htp!]
		\centering
		\includegraphics[width=0.8\textwidth]{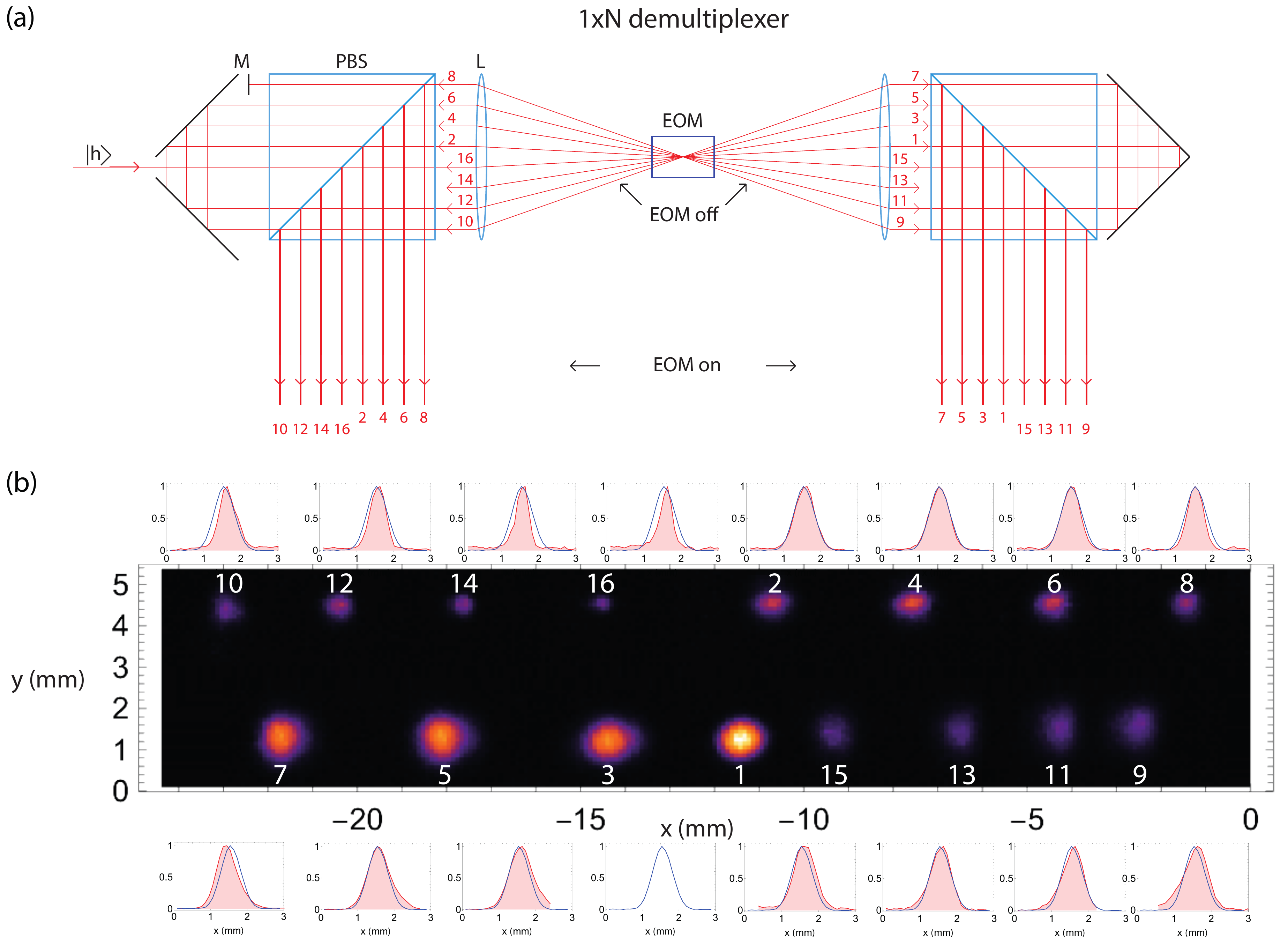}\vspace{0mm}
		\caption{\textbf{16-output demultiplexer.} (a) Same concept as described in the main text, but with a smaller path delay of about $1$~ns. Here, smaller beams of about $1$~mm diameter can propagate, and fit up to 16 modes using only standard 1-inch optics. (b) Image of all 16 outputs within 1-inch clear aperture, taken with laser light, and only passively allowing a small portion of the traveling beam to reflect from the PBSs. The small blue Gaussian fit corresponds to the transverse profile of the first output. All other 15 outputs (red areas) maintain a similar size.}
	\label{figSM:3}
	\end{figure*}	

\end{document}